\documentclass[preprint]{aastex}

\usepackage{epsfig}
\usepackage{graphicx}

\shorttitle{Speckle control with a PIAA}
\shortauthors{Martinache et al}

\begin{document}

\title{Speckle Control with a remapped-pupil PIAA-coronagraph}

\author{Frantz Martinache, Olivier Guyon, Christophe Clergeon \& Celia
Blain}
\affil{National Astronomical Observatory of Japan, Subaru Telescope,
  Hilo, HI 96720, USA}
\email{frantz@naoj.org}

\begin{abstract}
The PIAA is a now well demonstrated high contrast technique that uses
an intermediate remapping of the pupil for high contrast coronagraphy
(apodization), before restoring it to recover classical imaging
capabilities.
This paper presents the first demonstration of complete speckle
control loop with one such PIAA coronagraph. We show the presence of a
complete set of remapping optics (the so-called PIAA and matching
inverse PIAA) is transparent to the wavefront control
algorithm. Simple focal plane based wavefront control algorithms can
thus be employed, without the need to model remapping effects.
Using the Subaru Coronagraphic Extreme AO (SCExAO) instrument built
for the Subaru Telescope, we show that a complete PIAA-coronagraph is
compatible with a simple implementation of a speckle nulling
technique, and demonstrate the benefit of the PIAA for high contrast
imaging at small angular separation.
\end{abstract}


\section{Introduction}

Contrast limits for the direct imaging of extrasolar planets from
ground based adaptive optics (AO) observations are currently set by
the presence of static and slow-varying aberrations in the
optical path that leads to the science instrument
\citep{2003EAS.....8..233M}. 
These aberrations, due to the non-common path error between the
wavefront sensor and the science camera are responsible for the
presence of long lasting speckles in the image. Because extrasolar
planets are unresolved sources, it is difficult to discriminate them
among these speckles. One family of techniques, called differential
imaging, is aimed at calibrating out some of these static aberrations,
by using either sky rotation (Angular Differential Imaging, or ADI),
polarization (PDI), or wavelength dependence of the speckles (Spectral
Differential Imaging or SDI).
Of these, ADI \citep{2006ApJ...641..556M} seems very well adapted to
the problem of the detection of extrasolar planets, and has been
successful, most notably producing the image of the planetary system
around HR 8799 \citep{2008Sci...322.1348M}.
ADI uses the rotation of the sky that naturally happens while tracking
with an alt-azimuthal telescope around transit. The position of static
and slowly varying speckles, tied to the diffraction by the pupil,
remains stable over long timescales, while the image of planetary
companions rotates around the one of the host star.

The rotation of the field only leads to sufficient linear displacement
for angular separations of the order of one arcsecond. And in
practice, below 0.5 arcsecond, the performance of ADI quickly
degrades below the threshold where planets can be detected.

One way to complement ADI toward small angular separation, is to use a
deformable mirror (DM) to modulate speckles and introduce the
diversity that will distinguish them from genuine structures like
planets and lumps in disks. This type of technique is regularly used
for high contrast experiments \citep{2010PASP..122...71G}, and appears
as the technique of choice for a space borne mission dedicated to the
direct imaging of high contrast planets. This is also the approach we
propose to use for the detection of extrasolar planets at small
angular separation during ground-based adaptive optics (AO)
observations.
In that scope we are integrating and testing the Subaru Coronagraphic
Extreme AO (SCExAO) project, whose optics have been described by 
\citet{2009PASP..121.1232L} and \citet{2011SPIE.8151E..22M}.

In comparison with other extreme-AO projects
\citep{2008SPIE.7015E..31M, 2010lyot.confE..44B}, SCExAO implements an
aggressive PIAA-coronagraph using a remapping of the pupil
\citep{2005ApJ...622..744G, 2006ApJ...639.1129M}
optimized for high-contrast detection at small angular
separations (down to 1 $\lambda/D$).

Laboratory high-contrast experiments relying on PIAA have demonstrated
high contrast imaging capabilitty in the 2-4 $\lambda/D$ angular
separation range, and achieved raw contrast of $\sim10^{-7}$ and beyond
\citep{2010PASP..122...71G,2011SPIE.8151E...3K,2011SPIE.8151E...1B},
that are several orders of magnitude beyond what a ground based
instrument is expected to produce \citep{2005ApJ...629..592G}.
From a reasonably good starting point, these experiments can produce
high contrast images in a fairly small number of iterations, using
electric field conjugation (EFC) framework that relies on an acute
knowledge of the system's complex amplitude response matrix
\citep{2006PhDT........47G,2006ApJ...638..488B}.
Ultimately, these techniques seem implementable at the telescope
once an extreme AO system produces a continuous stream of stable
high-Strehl images. They remain for now limited to the pampered
environment of the laboratory.
At this stage of its development (what is refered to as SCExAO Phase
1), SCExAO does not include a fully functioning high order wavefront
sensor. The 32x32 DM it implements can be nevertheless used to
actively generate speckle diversity and supplement ADI at small
angular separations.
SCExAO phase 1 relies on iterative speckle nulling, to produce a dark
hole \citep{1995PASP..107..386M} in the field.

\section{Speckle control within a PIAA-coronagraph}
\label{sec:method}

The Phase Induced Amplitude Apodization (PIAA) coronagraph
\citep{2003A&A...404..379G} is a high-contrast imaging device that
enables the detection of faint sources down to angular separation
$\sim$ 1 $\lambda/D$.
A PIAA is made of a set of aspheric optics designed to apodize the
pupil by geometrically redistributing the light while preserving the
overall collimation of the beam for an on-axis source. Because it
alters the pupil, the point spread function (PSF) of one such system
is however no longer translation invariant.
On SCExAO, the impact of the PIAA on the pupil is quite dramatic as it
goes as far as filling the void left by the 30 \% central obscuration
of the Subaru Telescope pupil, to produce a apodized pupil better
suited for high contrast imaging. The impact for off-axis sources is
therefore also quite spectacular as it produces strongly elongated
pineapple-shaped PSFs beyond a few $\lambda/D$
\citep{2009PASP..121.1232L}.

\begin{figure}
\plotone{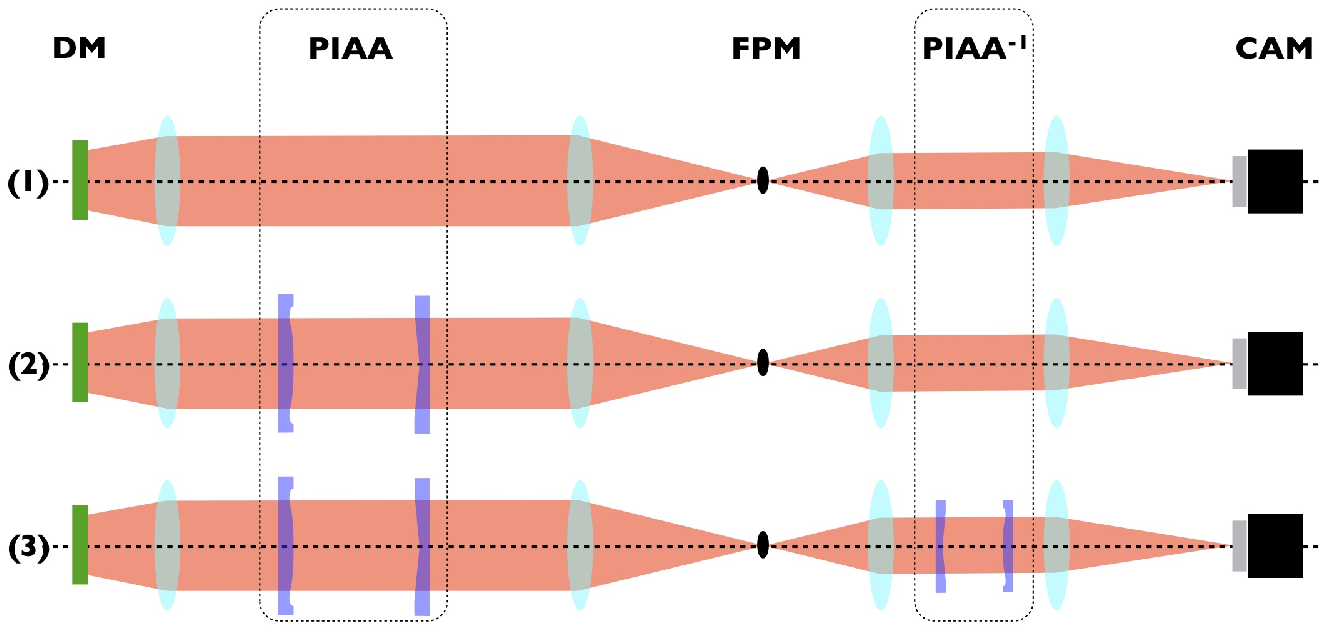}
\caption{Schematic representation three of the possible optical
  configurations of the SCExAO bench used for the results presented in
  this paper. From left to right, the main components are highlighted
  as follows: the deformable mirror (labeled DM) located at an
  intermediate   plane between pupil and image, the PIAA located in a
  collimated beam, the focal plane mask (labeled FPM) occulting the
  bright source   observed with the system, the inverse PIAA
  (labeled PIAA$^{-1}$), also located in a collimated beam, in a plane
  conjugated with the PIAA so as to remap the pupil back to what it is
  when it enters the system, and finally, the detector (labeled CAM).
  From top to bottom, the three configurations are: (1) no remapping
  optics, equivalent to a Lyot coronagraph without a Lyot stop, (2)
  with PIAA only, and (3) with PIAA and inverse PIAA (the full
  coronagraph).
}
\label{fig:configs} 
\end{figure}

While successful wavefront control experiments using PIAA have already
been reported \citep{2010PASP..122...71G, 2011SPIE.8151E...3K,
  2011SPIE.8151E...1B}, these experiments have all been using a DM
located downstream of the PIAA.
The outer working angle (OWA) of this type of control is imposed by
the total number of available actuators across one pupil diameter. By
placing the DM downstream of the remapping optics, the OWA is reduced
by a factor $\sim$ 3, due to the plate scale change induced by the
remapping \citep{2010PASP..122...71G}.
The resulting $\sim 5 \lambda/D$ OWA (since there are 32 actuators
across the pupil) would be a very serious limitation for any direct
imaging instrument.
Instead, the SCExAO project implements a complete PIAA-coronagraph,
with a DM located before any of the remapping optics
(cf. Fig. \ref{fig:configs}).
After the focal plane mask, a copy of the PIAA optics plugged
backwards (refered to as the inverse PIAA) is introduced to undo
the remapping of the pupil after the focal plane mask. This setup
allows to recover the wide field of view imaging capability of the
instrument \citep{2002A&A...391..379G, 2003A&A...404..379G}.

The SCExAO instrument is a flexible platform, installed after Subaru
Telescope's facility AO system, and designed to be used with the
coronagraphic imager HiCIAO \citep{2008SPIE.7014E..42H}, and increase
the size of the parameter space currently explored by the SEEDS survey
\citep{2009AIPC.1158...11T, 2010lyot.confE..15T}, toward small angular
separations.
The infrared arm of SCExAO alone \citep{2011SPIE.8151E..22M} supports
multiple optical configurations (cf. Fig. \ref{fig:configs}), from
straightfoward imager (with or without a focal plane mask), to a
complete PIAA-coronagraph.

\citet{2009PASP..121.1232L} have already demonstrated that the inverse
PIAA once aligned and conjugated to the PIAA, indeed restores wide
field of view imaging capability of the system, at least up to 20
$\lambda/D$, which is beyond the outer working angle defined by
the number of DM actuators across the instrument pupil. 
With this in mind, it seems reasonable to assume that the remapping
optics should therefore be completely transparent for the wavefront
control. 
The results presented in this paper demonstrate that indeed, with a
complete PIAA-coronagraph (including the inverse PIAA), a simple
wavefront control loop (speckle nulling) converges while being
oblivious to the two remappings of the pupil.

\subsection{Optical configurations of SCExAO}

To illustrate the impact of the remapping optics on the wavefront
control, we use three distinct optical configurations of the SCExAO
infrared arm, shown in Fig. \ref{fig:configs}. 
The reader will observe that in its current implementation, the 32x32
MEMS DM of SCExAO is not located in a pupil plane. This design was
chosen for the simplicity of the optical layout, as well as to
minimize the total number of reflections in the system. The design
however also has some drawbacks.
First, it requires to be somewhat conservative with the beam size
projected on the DM, to account for beam walk effects that would
otherwise significantly vignet the field. The DM also needs to be
horizontally tilted ($\sim$24$^{\circ}$), so the density of actuators is
not the same in horizontal and vertical directions. In practice,
instead of illuminating all 32 available actuators across one pupil
diameter, the beam spreads over a 27.2 x 24.8 actuator ellipse.
SCExAO can afford to have its DM away from the pupil since it is used
downstream from an AO system that not only stabilizes the pupil, but
also considerably reduces the total amount of aberrations the system
needs to correct for.

\begin{figure}
\plotone{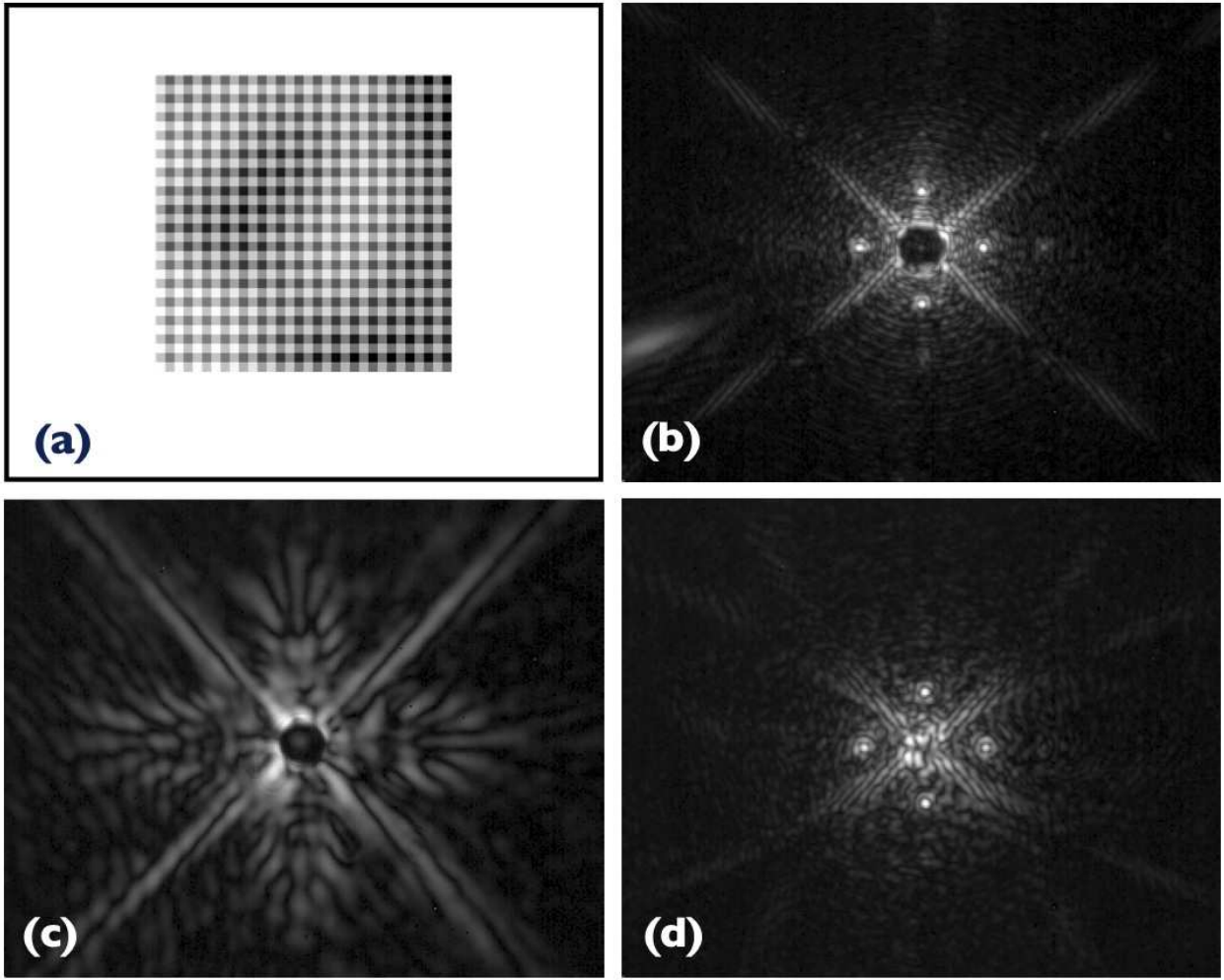}
\caption{
  Demonstration of the effect of remapping optics on the image
  recorded by SCExAO's science camera. Panel (a) shows the voltmap
  (ranging from 85 to 125 V) applied to the MEMS DM to produce the
  images that follow. Panels (b), (c) and (d) use a non-linear (square
  root) intensity scaling to highlight the faint diffraction
  structures surrounding the bright central area.
  Panel (b) shows the science camera image when the system is in
  configuration (1) (cf. Fig. \ref{fig:configs}). Panel (c) shows the
  dramatic effect of the PIAA on the focal plane image.
  Panel (d) shows the image obtained with the full coronagraph.
  Because the PIAA + inverse PIAA successfully restores the image of
  the off-axis speckles, it is expected that any type of focal-plane
  based wavefront control can ignore the presence of these remapping
  optics.
}
\label{fig:remap} 
\end{figure}

While the DM is always part of the optical train, both the PIAA and
the inverse PIAA are mounted on motorized stages that can swing in and
out of the beam with excellent repeatability, thus allowing for quick
alternance between three configurations: no remapping, PIAA only and
PIAA + inverse PIAA (cf. Fig. \ref{fig:configs}).
In a speckle control loop, the DM is used to introduce spatial
frequencies that destructively interfere with speckles induced by
wavefront aberrations from the input beam. The DM can also be used to
create speckles, and in turn, provide a very instructive demonstration
of how (cf. Fig. \ref{fig:remap}) aberrations propagate through the SCExAO
coronagraph.

Panel (a) of Fig. \ref{fig:remap} shows the 32x32 volt-map sent to the
DM for this experiment. In addition to the volt-map that has been
determined to produce the best output wavefront in the absence of
input aberrations (what will now be refered to as the DM flat-map),
two sine waves of amplitude 0.4 radians and maximal spatial frequency
(in the Nyquist sense) are added, along the horizontal and vertical
directions.
These additional sine functions on the DM create four bright speckles,
clearly visible in the focal plane visible on Panel (b) of
Fig. \ref{fig:remap}, in the absence of remapping optics.
The oversized focal plane mask effectively hides the brightest area,
but in the absence of Lyot stop, the diffraction by the spider arms
bearing the secondary mirror of the telescope are left.
Because the speckles result from the highest spatial frequency that
can be introduced by the DM, their locations in the image also
mark the edges of the region that can be controlled by the
DM. In the absence of amplitude defects, appropriate actuation of the
DM could clear the speckles over the entire 27.2 x 24.8 $\lambda/D$
box centered on the on-axis target.
In practice, because the beam includes amplitude as well as phase
defects, the DM can only operate over a half of this entire
region. For the results reported in this work, we arbitrarily chose to
work on the left-hand side of the field. Speckles located beyond 13.6
$\lambda/D$ along the horizontal axis and 12.4 $\lambda/D$ along the
vertical axis simply will not be affected by the the DM.
Note that with a 0.04 arcsecond diffraction limit in the H-band, 
12.4 $\lambda/D$ on the Subaru Telescope closely matches the range of
angular separation where ADI becomes usable
($12.4\times0.04\simeq0.5$ arcsecond).

\subsection{Impact of the remapping optics}

The introduction of the PIAA dramatically impacts the structure of
these image (cf. panel (c) of Fig. \ref{fig:remap}). 
While the diffraction spikes created by the spider remain, most of the
diffraction rings due to the sharp edge of the telescope pupil
disappear, and are more effectively hidden by the focal plane mask
whose size now better matches the on-axis PSF.
Off-axis, however, the bright diffraction-limited images of introduced
speckles are turned into complex pineapple-shaped aberrated
structures.
The benefit of the PIAA remapping will be better demonstrated when
comparing the performance of the speckle nulling loop with and without
using the remapping optics in Section \ref{sec:vmaps}.

The complete PIAA-coronagraph (configuration 3) includes after the
focal plane mask, a copy of the PIAA optics mounted backwards to
restore the geometry of the pupil back to what it was like before
entering the coronagraph. From geometric optics principles only, one
expects this restoration to be perfect, but diffraction and the
presence of the occulting mask in the focal plane will impact the
output wavefront. 
Panel (d) of Fig. \ref{fig:remap} shows the resulting image: the
bright off-axis speckles have been fully reconstructed by the inverse
PIAA, and have recovered their airy-disk shape. The dark disk left by
the focal plane mask that was visible in the other two images is no
longer obvious, as the disk is remapped by the inverse PIAA.

The PIAA + inverse PIAA combination successfully restores the image of
the off-axis speckles back into translation invariant copies of the
central PSF. For speckles that would be sufficiently far from the edge
of the focal plane mask, it is expected that any type of focal-plane
based wavefront control can ignore the presence of these remapping
optics. It is however not obvious for the speckles located nearby the
original footprint of the focal plane mask that they should behave as
simply. Section \ref{sec:results} will however show that speckle
control can be achieved within the entire control region of the DM.

\section{Speckle nulling results}
\label{sec:results}

\subsection{The PIAA-coronagraph is transparent to the speckle
  control}

To test the impact of the remapping (PIAA + inverse PIAA) optics for
a focal plane based wavefront control loop, we use a simple iterative
speckle nulling algorithm \citep{1995PASP..107..386M}, able to handle
up to a dozen speckles simultaneously.
This approach seems fairly inefficient in comparison with more
sophisticated electric field conjugation (EFC) approaches that can
achieve the same result in just a few iterations
\citep{2006PhDT........47G}. Speckle nulling was nevertheless chosen
because of its robustness and ease of implementation.

Indeed, complete EFC-based techniques rely on the knowledge of a
response matrix that relates the complex amplitude of the on-axis
source in the focal plane to the input wavefront. With an extreme AO
loop effectively stabilizing the wavefront entering the coronagraph
into something predictable, this matrix is likely to be quite stable,
and therefore reliable.
However in its first phase of deployment at the telescope, SCExAO does
not include the fast wavefront sensor. The simple speckle nulling
approach should, despite the presence of a dynamic atmospheric
component, calibrate out the wavefront features responsible for the
presence of static speckles in the images \citep{2010PASP..122...71G}.

Iterative speckle nulling works as follows: in a given image, up to $n$
speckles are identified and their positions marked, relative to the
central source. 
With a conventional imaging system, to each speckle position
corresponds a two-component $(x,y)$ spatial frequency on the DM, while
its brightness indicates the amplitude a$_0$ of this spatial frequency. 
The only real unknown is the phase $\varphi$ of each speckle, that can
take any value between 0 and 2$\pi$.
In the following four acquisitions, to each speckle is added a speckle
probe of same amplitude a$_0$, but each time with a different
phase: 0, $\pi$/2, $\pi$ and 3$\pi$/2. The intensity of the four speckles
resulting from the interference of the original speckle and the probes
is used to determine its true phase $\varphi_0$.
For the next image, also used as the input for the next iteration, a
spatial frequency of opposite phase $\varphi_0+\pi$ and of
amplitude $g \times a_0$, where $g$ is the loop gain (
$0.0 < g < 1.0$) is added so as to suppress the speckle.

\begin{figure}
\plotone{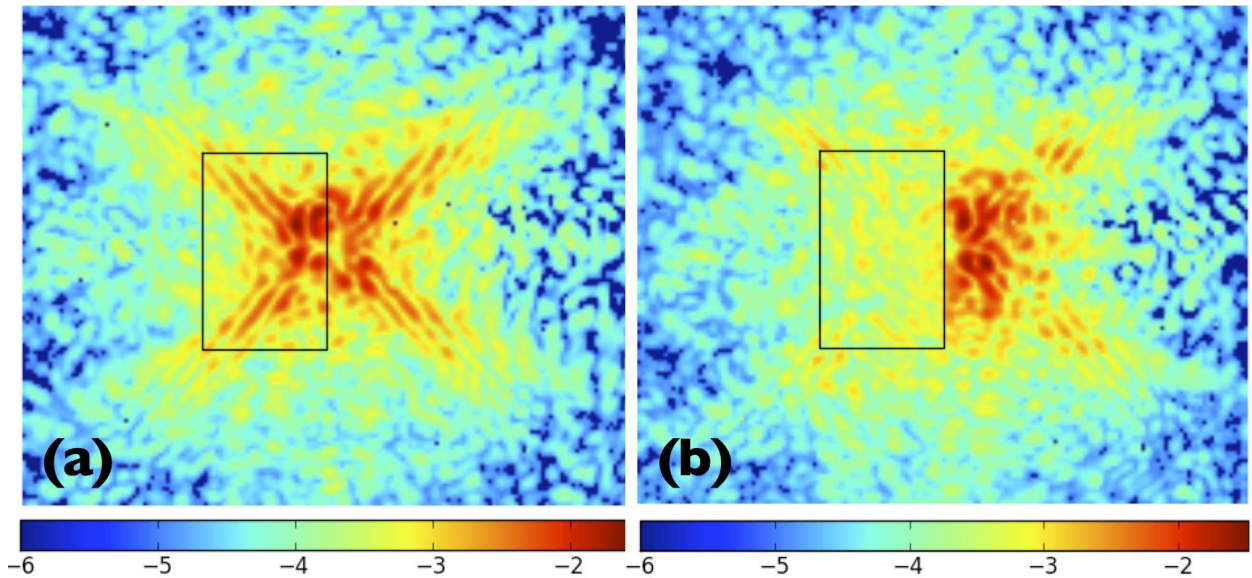}
\caption{
  Example of high contrast result achieavable with the SCExAO
  coronagraph using a simple speckle nulling control loop.
  Panel (a) shows the starting point of the loop, with the deformable
  mirror in its nominal flat-map configuration. Note that in addition
  to some low-spatial frequency aberrations (created by a static
  turbulence plate), most of the speckles present at the starting
  point are located along the diffraction spikes created by the spider
  arms of the telescope pupil.
  Panel (b) shows the result of about 50 speckle nulling iterations,
  working on up to 10 speckles at a time, effectively clearing a
  box-shaped region of speckles, from 0 to 14 $\lambda/D$ in the
  horizontal direction and within $\pm$ 14 $\lambda/D$ in the vertical
  direction.
}
\label{fig:spkn} 
\end{figure}

The result of a series of 50 such speckle nulling iterations is
presented in Fig. \ref{fig:spkn}. The starting point, shown in the
left panel, shows the structure of the speckles due to amplitude and
phase defects that filtered through the coronagraph. One will observe
that all ring like structures due to the sharp inner and outer edge of
the pupil have been erased from the image, showing that the PIAA -
focal plane mask - inverse PIAA combination achieves its
purposes. Although faint (the average contrast over the entire control
region is $2\times10^{-3}$), the most striking features can almost
entirely be attributed to the spider arms bearing the secondary.
After about 50 speckle nulling iterations (cf. right panel of Fig. 3),
the average contrast inside the control region is brought down to
$4\times10^{-4}$. The algorithm manages to suppress the diffraction
features due to the spider arms, along which the gain in raw contrast
is over $10^2$. Overall, the speckle nulling loop improves the
contrast by a factor 10 over the entire control region.

\subsection{The coronagraph relieves the wavefront control}
\label{sec:vmaps}

\begin{figure}
\plotone{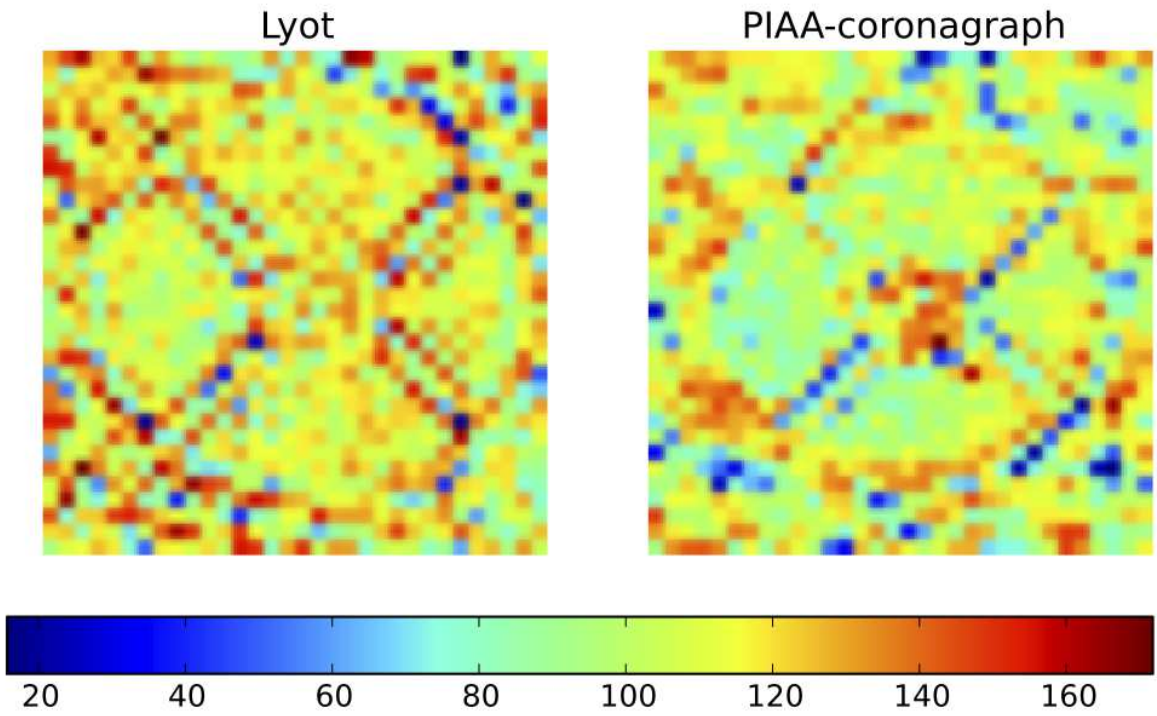}
\caption{
Speckle nulling voltmaps for configurations (1) Lyot-coronagraph and
(3) PIAA-coronagraph of SCExAO (cf. Fig. \ref{fig:configs}).}
\label{fig:vmap}
\end{figure}

Fig. \ref{fig:vmap} shows the DM voltmaps resulting from the
speckle nulling algorithm for the Lyot-coronagraph (left panel) and
the PIAA-coronagraph (right panel) configurations of SCExAO.
Despite being ignorant of the pupil geometry, the speckle nulling
produces a DM voltage induced phase pattern that resembles the
original telescope simulator pupil.

In addition to the spider vanes, in the Lyot configuration, the
algorithm attempts to cancel the diffraction rings induced by the
sharp inner and outer edge of the pupil. The corresponding voltmap
exhibits some sharp voltage changes from one actuator to its direct
neighbor near the projection of these edges on the DM.
Despite these sharp changes, this configuration fails to seriously
cancel the innermost rings that would require more range than the DM
can actually provide.
The PIAA-coronagraph voltmap on the other hand, exhibits smoother
features near the edge of the pupil, and only the effort to interfere
with the speckle due to the spider vanes are really obvious.
The coronagraph succeeds in relieving the wavefront control device by
effectively suppressing a major fraction of the diffraction in the
focal plane.

Note that the regions of the DM that obviously fall outside of the
pupil do not need to be actuated at all. A more sophisticated DM
control software should regularize the DM shape so as to preserve
stroke on the DM.

\subsection{Inner working angle}

\begin{figure}
\plotone{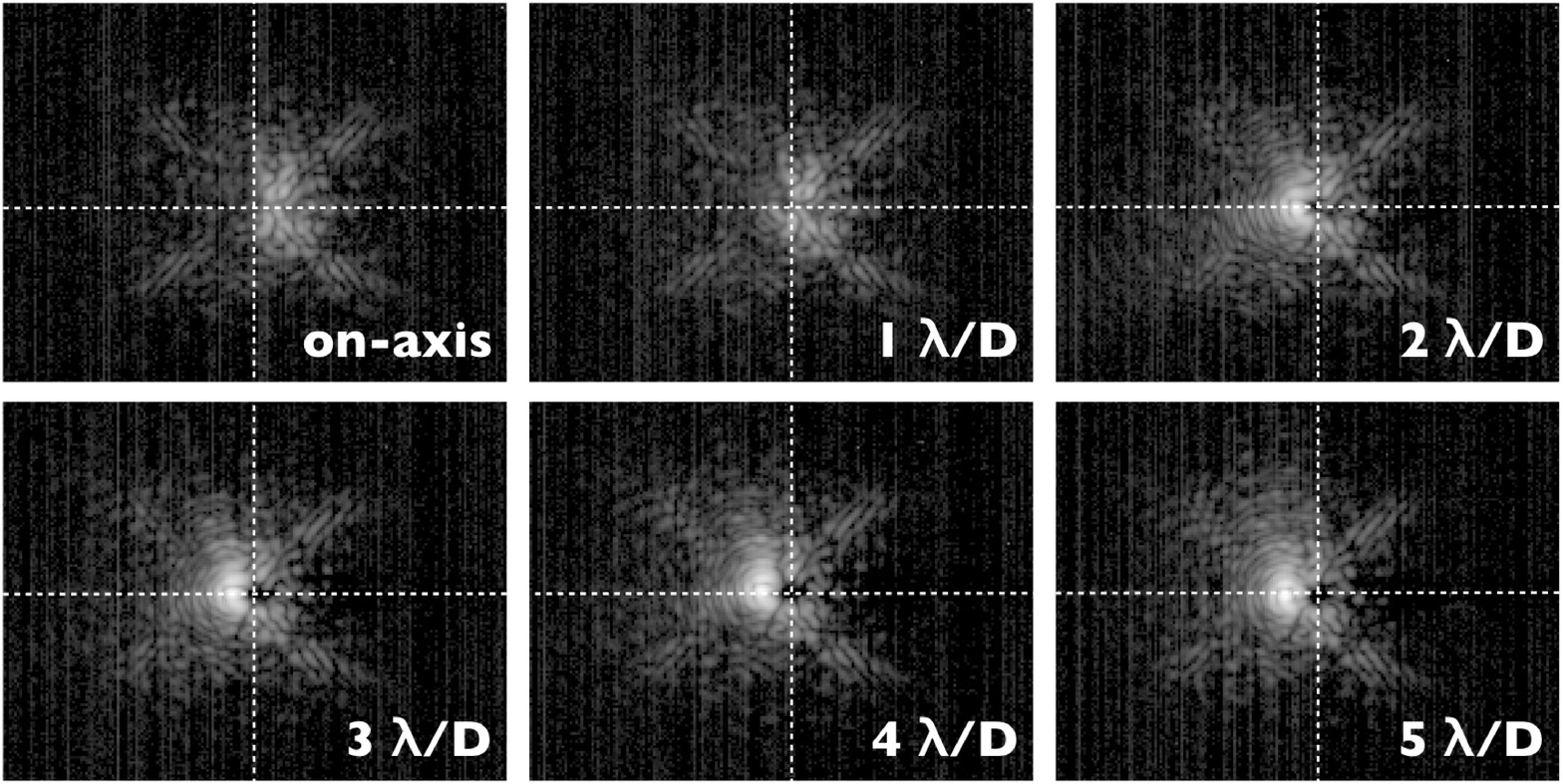}
\caption{
Evolution of the PSF as a function of off-axis position near the focal
plane mask from on-axis to 5 $\lambda/D$ off-axis. At 2 $\lambda/D$,
the core of the PSF has almost entirely cleared the focal plane mask,
and the diffraction rings associated with it become visible
again. Beyond 3 $\lambda/D$, the PSF morphology is essentially
translation invariant. Images use a common logarithmic intensity
scale to reveal both bright and faint features of the PSFs.
}
\label{fig:snaps}
\end{figure}

To further characterize the performance of the SCExAO coronagraph, and
estimate its inner working angle, we look at the evolution of the
throughput as a function of the position of the source relative to the
focal plane across the field. While applying the DM voltage map
resulting from a sequence of speckle nulling iterations, the source is
translated along the horizontal axis.
Fig. \ref{fig:snaps} shows six snapshots of the PSF as it moves across
the "dark hole" created by the speckle nulling loop for the first 5
$\lambda/D$ away from on-axis. Within 1 $\lambda/D$ and a little beyond,
ring-like structures remain erased by the coronagraph. At 2
$\lambda/D$, the core of the PSF clears the coronagraph and diffraction
rings re-appear in the image. Beyond 2 $\lambda/D$, the structure of
the PSF is essentially translation invariant, and only the shadow of
the focal plane mask betrays the presence of the coronagraph.

\begin{figure}
\plotone{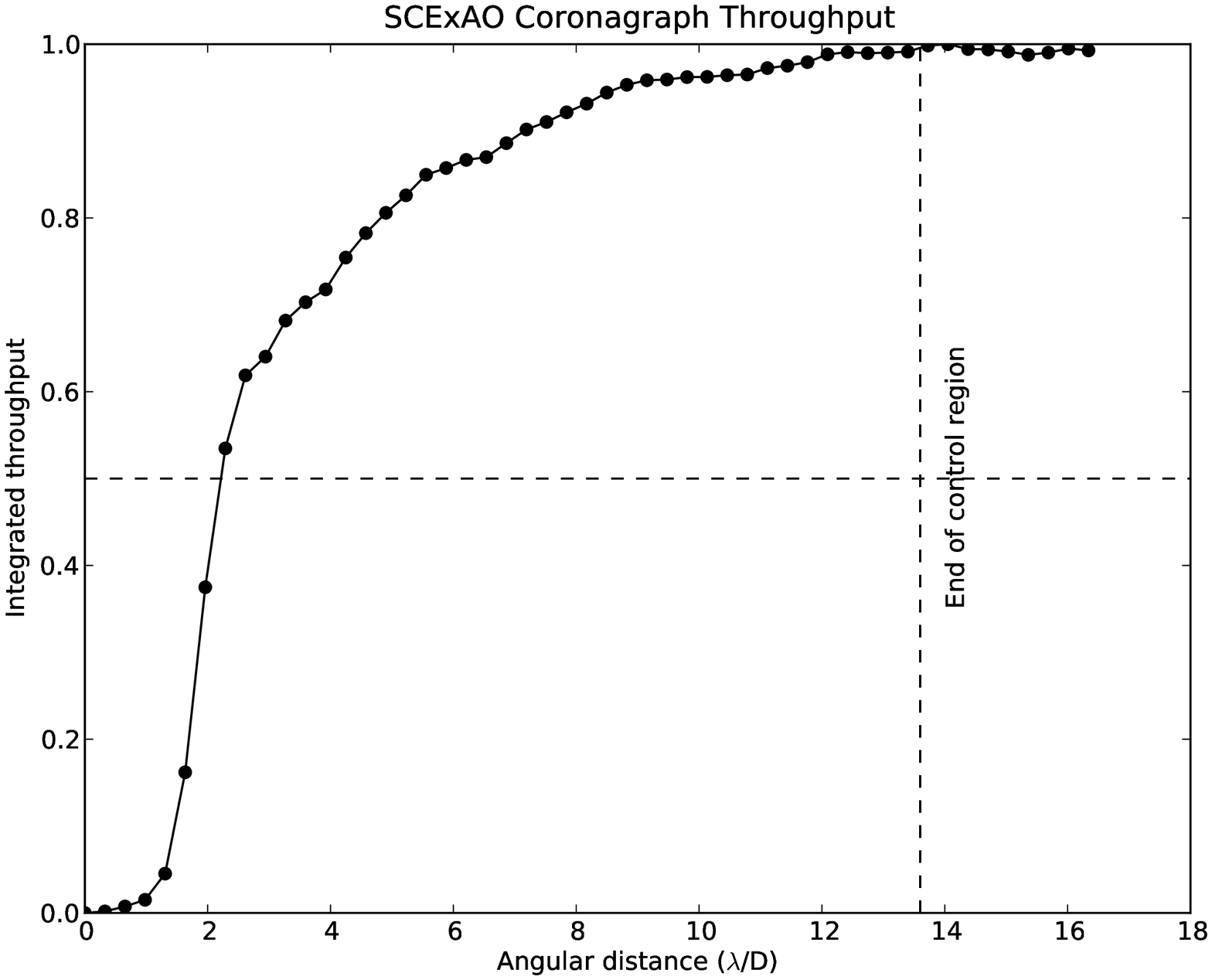}
\caption{
Throughput of the SCExAO coronagraph as a function of angular
separation. Using an iterative speckle nulling algorithm, the voltage
map that cancels speckles within the control region of the DM has been
identified. This curve shows how the throughput of the coronagraph
evolves as the source is progressively translated off-axis, across the
entire 14 $\lambda/D$ wide control region.
}
\label{fig:thrpt} 
\end{figure}

To complete this qualitative analysis of post coronagraph images,
Fig. \ref{fig:thrpt} shows a curve of the system throughput,
integrated over the left hand side of the field, as the sources is
moved across the entire 14 $\lambda/D$ control region, and slightly
beyond.
The 100 \% throughput reference is determined on-axis by removing the
focal plane mask.
Matching our qualitative description of snapshots in
Fig. \ref{fig:snaps}, we confirm that within 1 $\lambda/D$, the
throughput only varies at the percent level.
The most rapid variation of throughput happens around 2 $\lambda/D$,
where the PSF core clears the focal plane mask as shown in
Fig. \ref{fig:snaps}: the throughput quickly rises from a few percent
to 60 \%.

One convenient definition of the inner working angle (IWA) of a
coronagraph is the angular separation at which its transmission
reaches 50 \% \citep{2006ApJS..167...81G}. 
According to the measurements summarized in Fig. \ref{fig:thrpt},
SCExAO's current IWA is 2.2 $\lambda/D$. The focal plane mask currently
used is 40 \% oversized, compared to the 1.6 $\lambda/D$ IWA the PIAA
optics have been manufactured for. It will therefore be replaced
before SCExAO's next observing run.

\section{Conclusion}

We have, for the first time, demonstrated focal plane wavefront
control capability in a complete PIAA - focal plane mask - inverse
PIAA system.
Using a simple iterative speckle nulling algorithm, we successfully
produced a high contrast region in the field (the so-called dark
hole), and therefore confirmed that the complete set of remapping
optics (PIAA and inverse PIAA) are transparent to the focal plane
based wavefront control.
We have also confirmed that the outer working angle allowable by the
number of actuators of the deformable mirror is preserved during the
double remapping.
This implies that PIAA-type coronagraphs can be very well be used on
ground and space-based telescopes with no loss of outer working angle
and do not require complex wavefront control algorithms. 
High performance high efficiency coronagraphy therefore does not
translate into increased system complexity.

Note that while we tested PIAA with a large mask imposing a 2.2
$\lambda/D$ inner working angle), the conclusions of this paper will
hold for more IWA-optimized PIAA coronagraphs, including the PIAACMC
\citep{2010ApJS..190..220G} that exhibit IWA $< 1 \lambda/D$.
Combined with focal plane based wavefront control to what this paper
demonstrates, such aggressive designs make coronagraphy onboard small
telescopes ($\sim$2 meters) in space a relevant option for the direct
detection of extrasolar planets. They also offer the potential for
imaging reflected light planets at very small separation with the
forthcoming generation of extrely large telescopes.

One should also note that the moderate contrast level results
presented here were achieved in monochromatic light. Yet at the
$10^{-4}$ to $10^{-6}$ raw contrast level ground-based extreme AO
systems will give access to, speckle nulling remains a valid option
in polychromatic light.
The coronagraph currently in place in SCExAO does not take into
account the spider vanes in the pupil, while these are responsible for
most of the light observed in the final focal plane of the instrument.
Consequently, the wavefront control algorithm must use pupil phase
introduced by the deformable mirror to remove, over half of the focal
plane, light diffracted by the spider vanes, as shown in figure
\ref{fig:vmap}.
This approach is simply not suitable in a broad spectral band, and the
SCExAO coronagraph will be updated to a PIAACMC type coronagraph to
solve this issue.

Finally, while the results presented in this paper demonstrate the raw
contrast required for ground-based system, it is unclear to which
extent the approach adopted in this work is suitable for the
$\sim10^{-9}$ contrast level that future space-based missions hope to
reach in order to image and study Earth-like planets.
While remapping propagation effects specific to the PIAA coronagraph
can be included in the wavefront control algorithm
\citep{2011SPIE.8151E..12K}, it is unknown to which extent such
effects need to be accounted for. The approach presented in this paper
should be tested, both in simulations and in laboratory experiments,
at higher contrast that required for a ground-based system. This work
will be conducted over the next two years on existing PIAA coronagraph
testbeds.


\end{document}